\documentclass[fleqn,8pt]{wlscirep}
\newcommand{\ket}[1]{\vert #1 \rangle}
\newcommand{\bra}[1]{\langle #1 \vert}
\newcommand{\ketbra}[2]{\vert #1 \rangle \langle #2 \vert}

\usepackage{amsmath}
\usepackage{amssymb}
\usepackage{graphicx}
\usepackage{epstopdf}
\usepackage{todonotes}
\usepackage{color}
\usepackage{ulem}
\usepackage{tocloft} 
\usepackage[fleqn]{nccmath} 
\usepackage{lmodern}
\usepackage{float}
\title{Incoherent-mediator for quantum state transfer in the ultrastrong coupling regime}
\author[1,2,$\dag$]{F. A. C{\'a}rdenas-L{\'o}pez}
\author[1]{F. Albarr\'an-Arriagada}
\author[1]{G. Alvarado Barrios}
\author[1,2,*]{J. C. Retamal}
\author[1,+]{G. Romero}

\affil[1]{Departamento de F\'isica, Universidad de Santiago de Chile (USACH), 
	Avenida Ecuador 3493, 9170124, Santiago, Chile}
\affil[2]{Center for the Development of Nanoscience and Nanotechnology 9170124, Estaci\'on Central, Santiago, Chile}

\affil[$\dag$]{francisco.cardenas@usach.cl}
\affil[*]{juan.retamal@usach.cl}
\affil[+]{guillermo.romero@usach.cl}

\begin{abstract}
We study quantum state transfer between two qubits coupled to a common quantum bus that is constituted by an ultrastrong coupled light-matter system. By tuning both qubit frequencies on resonance with a forbidden transition in the mediating system, we demonstrate a high-fidelity swap operation even though the quantum bus is thermally populated. We discuss a possible physical implementation in a realistic circuit QED scheme that leads to the generalized Dicke model. This proposal may have applications on hot quantum information processing within the context of ultrastrong coupling regime of light-matter interaction. 
\end{abstract}

\begin{document}

\flushbottom
\maketitle
\keywords{Quantum state transfer, ultrastrong coupling regime, superconducting circuits}
%
%
\thispagestyle{empty}

\section*{Introduction}
The exchange of information between nodes of a quantum network is a necessary condition for large-scale quantum information processing (QIP) and networking. Many physical platforms have been proposed to implement high-fidelity quantum state transfer (QST) such as, coupled cavities \cite{PhysRevLett.78.3221,PhysRevLett.78.4293,PhysRevA.78.063805,PhysRevA.89.013853}, spin chains \cite{PhysRevLett.92.187902,PhysRevA.71.032312}, trapped ions \cite{PhysRevLett.74.4091,Haffner2008,RevModPhys.82.2313}, photonic lattices \cite{PhysRevA.87.012309}, among others. In the majority of cases, a necessary condition to carry out the transfer protocol is accessing to a highly controllable mediator. However, several protocols have been proposed to perform QST even though the mediator system is not controlled \cite{PhysRevLett.91.207901,PhysRevLett.106.040505}, or it is initialized in a thermally populated state \cite{PhysRevA.59.R2539,PhysRevLett.82.1971,PhysRevA.62.022311,Casanova2010,schuetz2016high}.

Likewise, the state-of-the-art quantum technologies has also proven useful for other quantum information tasks such as quantum computing \cite{Ladd:2010aa} and quantum simulations \cite{RevModPhys.86.153}. As important representatives of quantum devices we can mention superconducting circuits \cite{Clarke:2008aa} and circuit quantum electrodynamics (QED) \cite{PhysRevA.69.062320,Nature_431,Chiorescu2004,Nature_451}. These technologies have also pushed forward to achieve the ultrastrong coupling (USC) \cite{USCphase,Niemczyk2010,PhysRevLett.105.237001,Forn_2016,andersen2016USCTransmon,Forn2016,PhysRevB.93.214501} and deep strong coupling (DSC) \cite{PhysRevLett.105.263603,Yoshihara_2016} regimes of light-matter interaction, where the coupling strength becomes comparable to or larger than the frequencies of the cavity mode and two-level system. In this case, the light-matter coupling is well described by the quantum Rabi model (QRM) \cite{Rabi1,Braak2010} and features a discrete parity symmetry. It has been proven that the above symmetry may be useful for quantum information tasks in the USC regime \cite{PhysRevLett.107.190402,PhysRevLett.108.120501,srep08621,PhysRevB.91.064503,PhysRevA.94.012328}.

\begin{figure}[t]
\centering
\includegraphics[width=10cm,height=10cm,keepaspectratio]{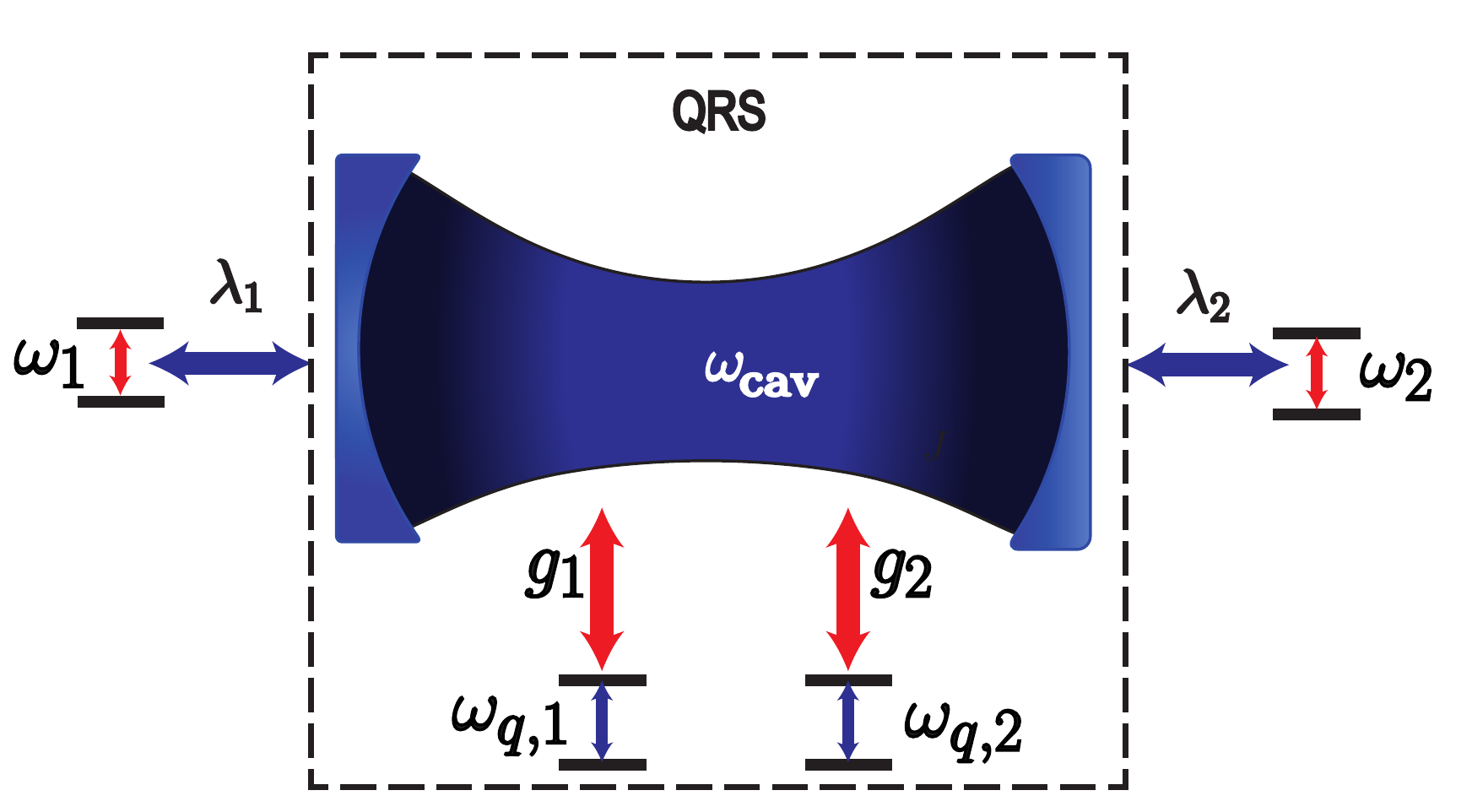}
\caption{\textbf{Schematic representation of the model}. Two qubits with frequency gaps $\omega_1$ and $\omega_2$ are coupled to a QRS via dipolar coupling with strengths $\lambda_1$ and $\lambda_2$, respectively. The QRS is constituted by a pair of two-level systems ultrastrongly coupled to a single cavity mode of frequency $\omega_{\rm cav}$.}
\label{fig:model}
\end{figure}

We propose a protocol for performing high-fidelity QST between qubits coupled to a common mediator constituted by a two-qubit quantum Rabi system (QRS) \cite{TQRM1,TQRM2}, see Fig.\ref{fig:model}. The QST relies on the tuning of qubit frequencies on resonance with a forbidden transition of the QRS, provided by the selection rules imposed by its parity symmetry. We demonstrate that high-fidelity QST occurs even though the QRS is thermally populated and the whole system experiences loss mechanisms. We also discuss a possible physical implementation of our QST protocol for a realistic circuit QED scheme that leads to the generalized Dicke model \cite{Phys.Rev.Lett.114. 143601} as mediating system. This proposal may bring a renewed interest on hot quantum computing \cite{PhysRevA.59.R2539,PhysRevLett.82.1971,PhysRevA.62.022311} within state-of-art light-matter interaction in the USC regime.

\section*{The model} 
Our proposal for quantum state transfer is schematically shown in Fig.~\ref{fig:model}. We consider a pair of two-level systems with transition frequencies $\omega_{q,i}$ ($i=1,2$), ultrastrongly coupled to a single cavity mode of frequency $\omega_{\rm{cav}}$. This situation is described by the two-qubit quantum Rabi model \cite{TQRM1,TQRM2}

\begin{ceqn}
\begin{eqnarray}
\label{QRS}
H_{\rm{QRS}} &=& \hbar \omega_{\rm{cav}}a^{\dag}a + \sum_{i=1}^{N=2}\frac{\hbar\omega_{q,i}}{2}\sigma_{i}^{z}+ \hbar g_{i}\sigma_{i}^{x}(a^{\dag} + a).
\end{eqnarray}
\end{ceqn}
Here, $a(a^{\dag})$ is the annihilation (creation) bosonic operator for the cavity mode. The operators $\sigma^{z}_{i}$ and $\sigma^{x}_{i}$ stand for Pauli matrices describing each two-level system. Also, $\omega_{\rm cav}$, $\omega_{q,i}$, and $g_{i}$, are the cavity frequency, $i$th qubit frequency, and $i$th qubit-cavity coupling strength, respectively. Furthermore, two additional qubits with frequency gaps $\omega_{n}$ ($n=1,2$), are strongly coupled to the QRS through the cavity mode with coupling strengths $\lambda_n$. The Hamiltonian for the whole system depicted in Fig.~\ref{fig:model} reads
\begin{ceqn}
\begin{eqnarray}
\label{model}
H = H_{\rm{QRS}} + \sum_{j=1}^{N=2}\frac{\hbar \omega_j}{2}\tau_{j}^{z} + \hbar\lambda_{j}\tau_{j}^{x}(a^{\dag}+a),
\end{eqnarray}
\end{ceqn}
where $\tau_{j}^{x,z}$ are Pauli matrices associated with the additional qubits. In what follows, we will discuss about the parity symmetry of the two-qubit quantum Rabi model (\ref{QRS}), and their corresponding selection rules.

\section*{Selection rules in the two-qubit quantum Rabi model}
An important result in quantum physics are the selection rules imposed by the electric and magnetic dipole transitions. Similarly, the parity symmetry ($\mathbb{Z}_{2}$) of the Hamiltonian (\ref{QRS}) imposes selection rules for state transitions. The $\mathbb{Z}_{2}$ symmetry can be seen if we replace $\sigma^{x}_i\to -\sigma^{x}_i$ and $(a^{\dag}+a)\to -(a^{\dag}+a)$ in the Hamiltonian~(\ref{QRS}) such that it remains unchanged. In other words, this symmetry implies the existence of a parity operator $\mathcal{P}$ that commutes with the Hamiltonian, $[H_{\rm QRS},\mathcal{P}]=0$. In this way both operators can be simultaneously diagonalized in a basis $\{\ket{\psi_j}\}^{\infty}_{j=0}$. In particular, for the two-qubit quantum Rabi model Eq.~(\ref{QRS}), the parity operator reads $\mathcal{P} = \sigma_{1}^{z}\otimes\sigma_{2}^{z}\otimes e^{i\pi a^\dag a}$, such that $\mathcal{P}|\psi_{j}\rangle = p |\psi_j\rangle$, with $p=\pm 1$, and $H_{\rm QRS}\ket{\psi_j}=\hbar\nu_{j}\ket{\psi_j}$, where $\nu_j$ is the $j$th eigenfrequency. 

\begin{figure}[t]
\centering 
\includegraphics[width=10cm,height=10cm,keepaspectratio]{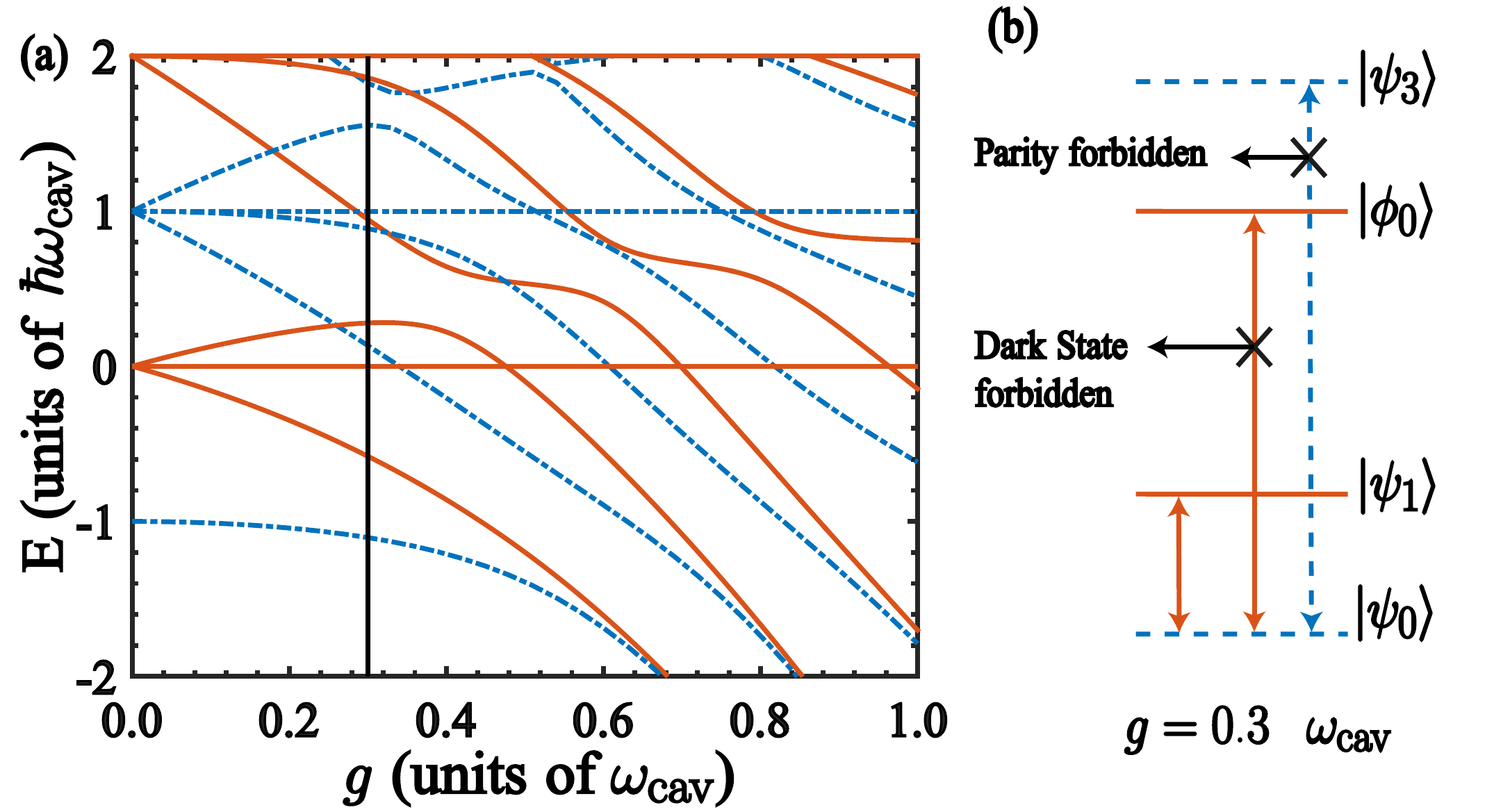}
\caption{\textbf{Energy Spectrum of the two-qubit quantum Rabi system}. (a) Energy spectrum of the Hamiltonian (\ref{QRS}) with parameters $\omega_{q,1} = \omega_{q,2} = \omega_{\rm{cav}}$, as a function of the coupling strength $g$. Blue (dot-dashed) lines stand for states with parity $p=+1$. Red (continuous) lines stand for states with parity $p=-1$. The straigth lines in the spectrum stand for dark states $\ket{\phi_N}$. (b) Schematics of the level structure at $g=0.3~\omega_{\rm cav}$ corresponding to the vertical solid line in (a). It is shown the allowed and forbidden transitions ruled by the parity symmetry of the two-qubit quantum Rabi model.}
\label{fig:spectrum}
\end{figure}

The selection rules associated with the parity symmetry appear when considering a cavity-like driving, proportional to $a^{\dag}+a$, or qubit-like driving $\propto \sigma^x$ or $\sigma^z$ \cite{Forn2016,PhysRevA.94.012328}. We are interested in the former case since each qubit in Fig.~\ref{fig:model} couples to the QRS via the field quadrature $X=a^{\dag}+a$. It is noteworthy that for the single-qubit quantum Rabi model \cite{Braak2010}, it is possible to demonstrate that matrix elements $q_{jk}=\bra{\varphi_j}(a^{\dag}+a)\ket{\varphi_k}$ are different from zero if states $\ket{\varphi_j}$ and $\ket{\varphi_k}$ belong to different parity subspaces, whereas $q_{jk}=0$ for states with same parity \cite{tomography}. In addition, for the two-qubit quantum Rabi model we need to take into account additional features. In this case, if one considers identical qubits and coupling strengths in the Hamiltonian~(\ref{QRS}), the spectrum features an invariant subspace formed by tensor products of pseudo spin and Fock states $\{|\downarrow,\uparrow,N\rangle,|\uparrow,\downarrow,N\rangle\}$, whose eigenstates are
\begin{ceqn}
\begin{eqnarray}
\ket{\phi_N} = \frac{1}{\sqrt{2}}\big(|\downarrow\uparrow\rangle-|\uparrow\downarrow\rangle\big)|N\rangle.
\label{DarkS}
\end{eqnarray}
\end{ceqn}
This is a dark state (DS) where the spin singlet is decoupled from the cavity mode \cite{1751-8121-47-13-135306,DARKQRM,peng2016dark}. Figure \ref{fig:spectrum}\textcolor{blue}{(a)} shows the energy spectrum of the Hamiltonian~(\ref{QRS}) for identical qubits ($\omega_{q,1}=\omega_{q,2}$) as a function of the coupling strength $g_1=g_2=g$. Blue (dot-dashed) lines stand for states with parity $p=+1$, while red (continuous) for states with parity $p=-1$. The dark states~(\ref{DarkS}) appear with constant energies whose gaps correspond to the cavity mode frequency.
The selection rules in the two-qubit quantum Rabi system need to take into account the emergence of dark states. In this case, each DS has definite parity as shown in Fig. \ref{fig:spectrum}\textcolor{blue}{(a)}; however, the matrix elements between a DS and remaining states are null, $X_{\phi_Nk}=\bra{\phi_N}(a^\dag+a)\ket{\psi_k}=0$, since states $\ket{\psi_k}$ can be written as linear superpositions of products of symmetric states for the pseudo spins and Fock states, that is, $\ket{\psi_k}=\sum^{\infty}_{N=0}\sqrt{N!}\{a_N(\ket{\uparrow\uparrow}\pm(-1)^N\ket{\downarrow\downarrow})+b_N(\ket{\uparrow\downarrow}\pm(-1)^N\ket{\downarrow\uparrow})\}\ket{N}$, where $\pm$ stands for parity $p=\pm1$ \cite{Duan2015,TQRabis_spectrum1}. In the light of the above forbidden transitions, we will show that they become a key feature for our quantum state transfer protocol.
\begin{figure}[t]
\centering 
\includegraphics[width=10cm,height=10cm,keepaspectratio]{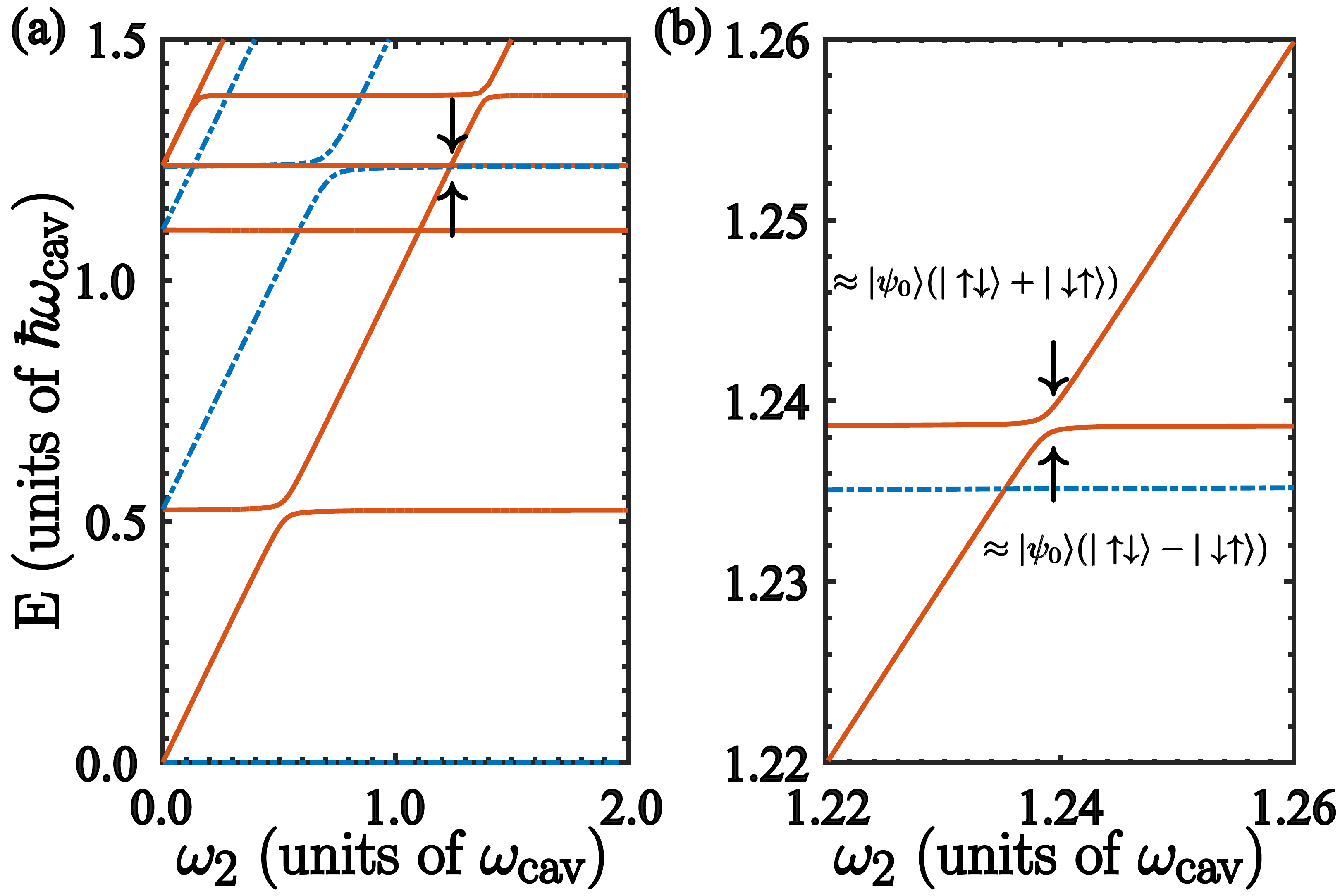}
\caption{\textbf{Spectrum of the complete model}.  (a) Energy differences from the spectrum of Hamiltonian (\ref{model}), as a function of the rightmost qubit frequency $\omega_{2}$. Blue (dot-dashed) lines stand for states with parity $p=+1$, while red (continuous) lines stand for states with parity $p=-1$. At frequency $\omega_{2}=\omega_{1}$ an avoided energy crossing appears. (b) Enlarged view of the energy spectrum shown in (a) around the region $\omega_{2}=\omega_{1}=\nu_{3}-\nu_{0}$ . The numerical calculation was performed with the same parameters used in Fig. \ref{fig:spectrum}.}
\label{fig:spectrumH}
\end{figure}

\section*{Parity assisted excitation transfer} 
We study the QST between two qubits that are coupled to a common QRS system [cf. Fig.~\ref{fig:model}]. We focus on the situation where frequencies of the leftmost and rightmost  qubits are resonant with a forbidden transition of the QRS. For instance, if we consider the case $g=0.3~\omega_{\rm cav}$ as denoted by the vertical solid line in Fig.~\ref{fig:spectrum}\textcolor{blue}{(a)}, we choose $\omega_{1}=\omega_2=\nu_{3}-\nu_{0}$. It is clear that the matrix element $X_{03}=\bra{\psi_0}a+a^{\dag}\ket{\psi_3}=0$ because states $\ket{\psi_0}$ and $\ket{\psi_3}$ have parity $p=+1$. Also, the matrix element $X_{02}=0$ since $\ket{\psi_2}=\ket{\phi_0}$ is a dark state. The allowed and forbidden transitions for the lowest energies of the Hamiltonian (\ref{QRS}), for $g/\omega_{\rm cav}=0.3$, are schematically shown in Fig.~\ref{fig:spectrum}\textcolor{blue}{(b)}. 

In view of the above conditions, and considering the QRS initialized in its ground state, we realize that the only transmission channel that would participate in a QST protocol between qubits corresponds to the first excited state of the QRS, since the matrix element $X_{01}\ne 0$, as $\ket{\psi_0}$ and $\ket{\psi_1}$ have opposite parity. However, both qubits are far-off-resonance with respect to this allowed transition. In this case, one can demonstrate that the QST occurs in a second-order process, as qubits interact dispersively with the QRS resulting in an effective qubit-qubit interaction. The latter can be seen in the spectrum of the Hamiltonian (\ref{model}), as depicted in Fig.~\ref{fig:spectrumH}\textcolor{blue}{(a)}. Specifically, around the region $E/\hbar\omega_{\rm cav}\approx 1.2386$ appears an avoided level crossing, enlarged in Fig.~\ref{fig:spectrumH}\textcolor{blue}{(b)}, where states $\ket{\psi_0}\ket{\uparrow\downarrow}$ and $\ket{\psi_0}\ket{\downarrow\uparrow}$ hybridize to form maximally entangled states well approximated by $\ket{\psi_0}(\ket{\uparrow\downarrow}+\ket{\downarrow\uparrow})/\sqrt{2}$ and $\ket{\psi_0}(\ket{\uparrow\downarrow}-\ket{\downarrow\uparrow})/\sqrt{2}$. This hybridization resembles the two-body interaction mediated by a single-qubit quantum Rabi model~\cite{kyaw2016polariton}. 

The effective qubit-qubit interaction can be obtained from the total Hamiltonian (\ref{model}) via a dispersive treatment beyond rotating-wave approximation \cite{kyaw2016polariton,PhysRevA.80.033846}, see the supplemental material for a detailed demonstration. In this case, the effective qubit-qubit interaction reads 
\begin{ceqn}
\begin{eqnarray}
\label{Eff_model}
H_{\rm{eff}} &=& H_{0} + \frac{\hbar}{2}|\chi_{10}|^2 \mathbf{Z}_{p}\otimes\mathbf{S}_{12},
\end{eqnarray}
\end{ceqn}
where 
\begin{ceqn}
\begin{eqnarray}
|\chi_{10}|^2 &=& |\langle \psi_0|(a^\dag + a)|\psi_1\rangle|^2\nonumber\\
\mathbf{Z}_p &=& |\psi_1\rangle\langle \psi_1| -  | \psi_0\rangle\langle\psi_0|\nonumber\\
\mathbf{S}_{12} &=& \lambda_1\lambda_2\Bigg(\frac{1}{\mu_{10}^{1}}  + \frac{1}{\mu_{10}^{2}}- \frac{1}{\Delta_{10}^{1}} - \frac{1}{\Delta_{10}^{2}}\Bigg)\tau^{x}_{1}\tau^{x}_{2} + 2\sum_{j=1}^{2}\lambda_{j}^2\Bigg(\frac{\tau^{+}_{j}\tau^{-}_{j}}{\Delta^{j}_{10}} - \frac{\tau^{-}_{j}\tau^{+}_{j}}{\mu^{j}_{10}}\Bigg)\nonumber,
\end{eqnarray}
\end{ceqn}
and $\tau_n^{\pm}=(\tau_j^{x}\pm i\tau_j^{y})/2$. The detunings are defined as $\Delta_{10}^{j} = \omega_{j} - \nu_{10}$ and $\mu_{10}^{j} = \omega_{j} + \nu_{10}$, where $\omega_{j}$ corresponds to the $j$th qubit frequency that interacts with the QRS, see Fig.~\ref{fig:model}, and $\nu_{10}$ stands for the frequency difference between the two lowest energy states of the QRS. 
\begin{figure}[t]
	\centering
	\includegraphics[width=10cm,height=10cm,keepaspectratio]{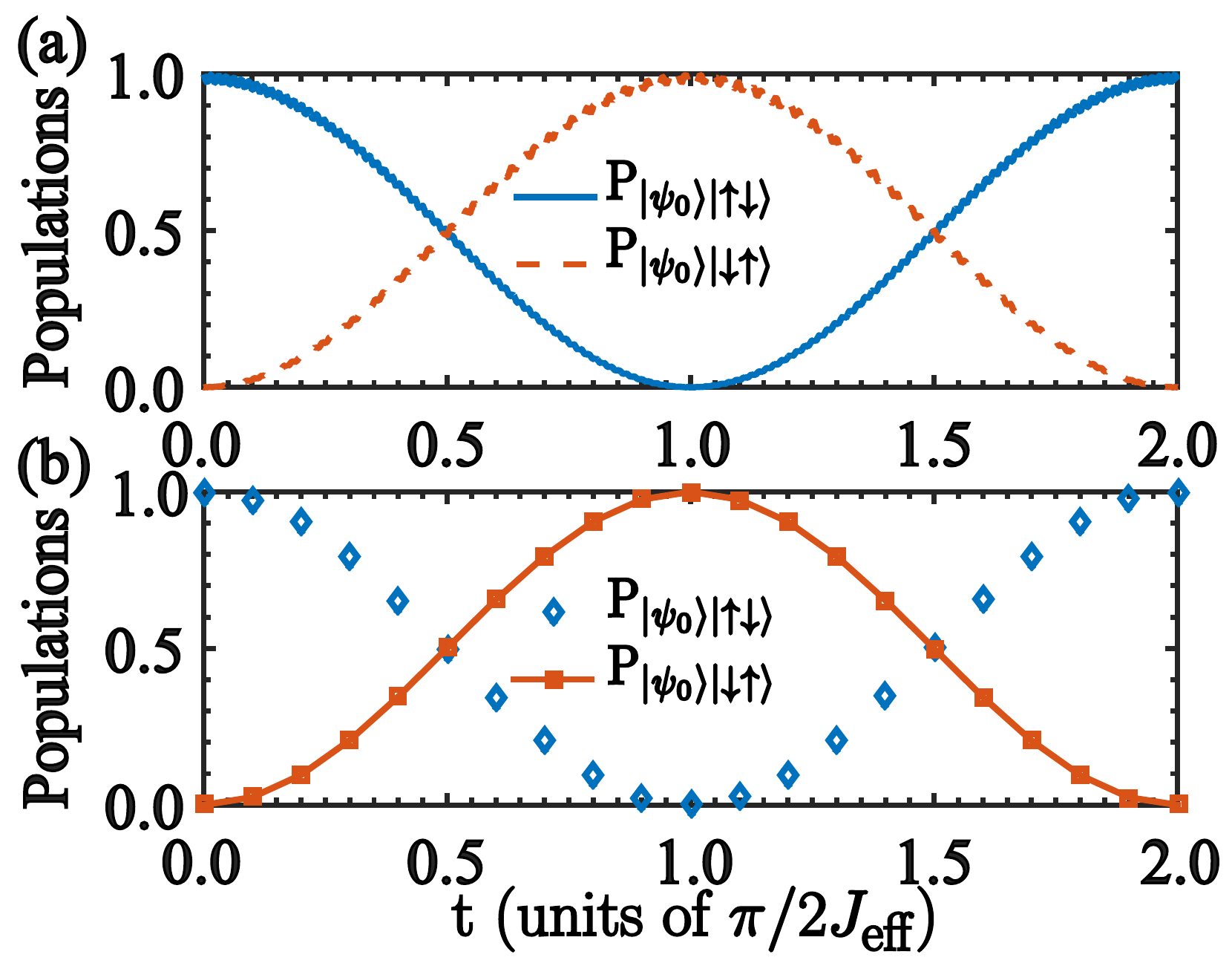}
	\caption{\textbf{Population evolution}. (a)	Population inversion of the states $|\psi_0\rangle|\uparrow\downarrow\rangle$ and $|\psi_0\rangle|\downarrow\uparrow\rangle$, numerically calculated from the {\it ab initio} model (\ref{model}). (b) Population inversion calculated from the effective Hamiltonian (\ref{Eff_model}). These numerical calculations have been performed with parameters, $\omega_{q,i} = \omega_{\rm{cav}}$, $g=0.3~\omega_{\rm{cav}}$, $\omega_{1}=\omega_2=\nu_{3}-\nu_{0}$ and $\lambda_{1}=\lambda_{2} = 0.02~\omega_{\rm{cav}}$.}
	\label{fig:effective}
\end{figure}

Figure \ref{fig:effective}\textcolor{blue}{(a)} shows the population inversion between states $\ket{\psi_0}\ket{\!\uparrow\downarrow}$ and $\ket{\psi_0}\ket{\!\downarrow\uparrow}$ calculated from the full Hamiltonian (\ref{model}), and from the dispersive Hamiltonian (\ref{Eff_model}), see Fig.~\ref{fig:effective}\textcolor{blue}{(b)}. In both cases, the qubit-qubit exchange occurs within a time scale proportional to the inverse of the effective coupling constant $J_{\rm{eff}}=|\chi_{10}|^2\lambda_{1}\lambda_{2}(1/\mu_{10}^{1}  + 1/\mu_{10}^{2}- 1/\Delta_{10}^{1} - 1/\Delta_{10}^{2})$. For parameters, $\omega_{q,i} = \omega_{\rm{cav}}$, $g=0.3~\omega_{\rm{cav}}$, $\omega_{1}=\omega_2=\nu_{3}-\nu_{0}$ and $\lambda_{1}=\lambda_{2} = 0.02~\omega_{\rm{cav}}$, we obtain $|\chi_{10}| = 1.0325$, $\mu_{10}^{1} = \mu_{10}^{2} =1.7632 ~\omega_{\rm{cav}}$ and $\Delta_{10}^{1} =\Delta_{10}^{2}=0.7141 ~\omega_{\rm{cav}}$. These values lead to an effective coupling strength $2J_{\rm eff}\approx 0.0011~\omega_{\rm{cav}}$. In addition, if we consider typical values for microwave cavities such as $\omega_{\rm cav}=2\pi \times 8.13$ GHz \cite{Forn2016}, the excitation transfer happens within a time scale of about $t =\pi/2J_{\rm{eff}}\approx 56~[{\rm ns}]$.

\begin{figure}[t]
\centering
\includegraphics[width=10cm,height=10cm,keepaspectratio]{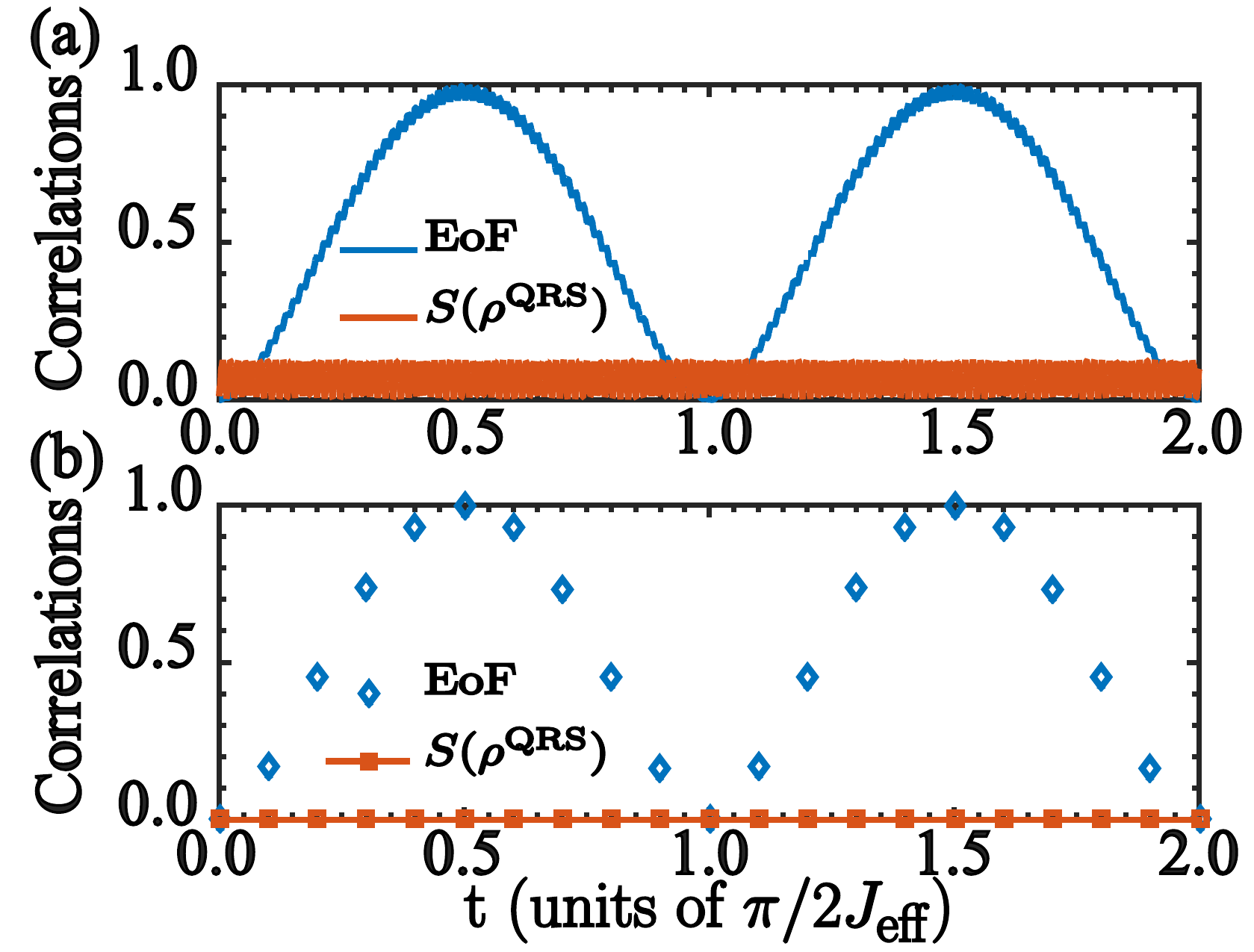}
\caption{\textbf{Correlation evolution}. (a) Entanglement dynamics for the reduced system composed by the leftmost and rightmost qubits, and the von Neumann entropy $S(\rho^{\rm{QRS}})$ for the reduced QRS system, numerically calculated from the {\it ab initio} model (\ref{model}). (b) Entanglement of formation and von Neumann entropy numerically calculated from the effective Hamiltonian (\ref{Eff_model}). We have used the parameters from Fig.~\ref{fig:effective}.}
\label{fig4}
\end{figure}

Additionally, we also study quantum correlations in our system. Figure \ref{fig4} shows the Entanglement of Formation (EoF) for the subsystem composed by the leftmost and rightmost qubits, and the von Neumann entropy, $S(\rho)$, for the reduced density matrix of the QRS. $S(\rho)$ shows a negligible correlation between the bipartition composed by QRS and the additional qubits. This behavior can be predicted from the effective Hamiltonian (\ref{Eff_model}), which shows an effective qubit-qubit interaction while it is diagonal in the QRS basis. Therefore, the QRS will not evolve and the quantum correlations between it and additional qubits is negligible. At the same time, the EoF between the leftmost and rightmost qubits has an oscillating behavior whose minimum value is reached when the excitation transfer has been completed.

\section*{Incoherent mediator for quantum state transfer} 
Notably, the above described mechanism allows us for quantum state transfer between qubits even though the QRS is initially prepared in a thermal state at finite temperature. Let us describe how our system governed by Eq.~(\ref{model}) is initialized in a thermally populated state. Since the frequencies of the additional qubits are larger than the lowest energy transition of the QRS, we should expect that thermal population associated with both qubits are concentrated only in their ground states. In order to ascertain the latter, we compute the fidelity $\mathcal{F}$ between the Gibbs state obtained from the Hamiltonian (\ref{model}), $\rho_{\rm Gibbs}=e^{(-H/k_BT)}/Z$, where $Z=\sum_m{\rm exp}(-\hbar \epsilon_m/k_BT)$ is the partition function, and the probe state defined by $\rho_{\rm{p}}=\rho_{\rm{th}}^{\rm{\small{QRS}}}\otimes |\downarrow\downarrow\rangle\langle\downarrow\downarrow|$, where $\rho_{\rm{th}}^{\rm{\small{QRS}}}$ is the Gibbs state for the QRS. Notice that $\hbar\epsilon_j$ corresponds to the $j$th eigenvalue of the Hamiltonian~(\ref{model}), $H\ket{m}=\hbar\epsilon_m\ket{m}$. Taking parameters from Fig. \ref{fig:effective} and $T=100$~mK, we obtain the fidelity $\mathcal{F}={\rm tr}(\rho_{\rm Gibbs}\rho_{\rm p})=0.9951$. Therefore, the Gibbs state is a tensor product between the additional qubits in their ground states and the QRS in an thermal state.
\begin{figure}[t!]
	\centering
	\includegraphics[width=10cm,height=10cm,keepaspectratio]{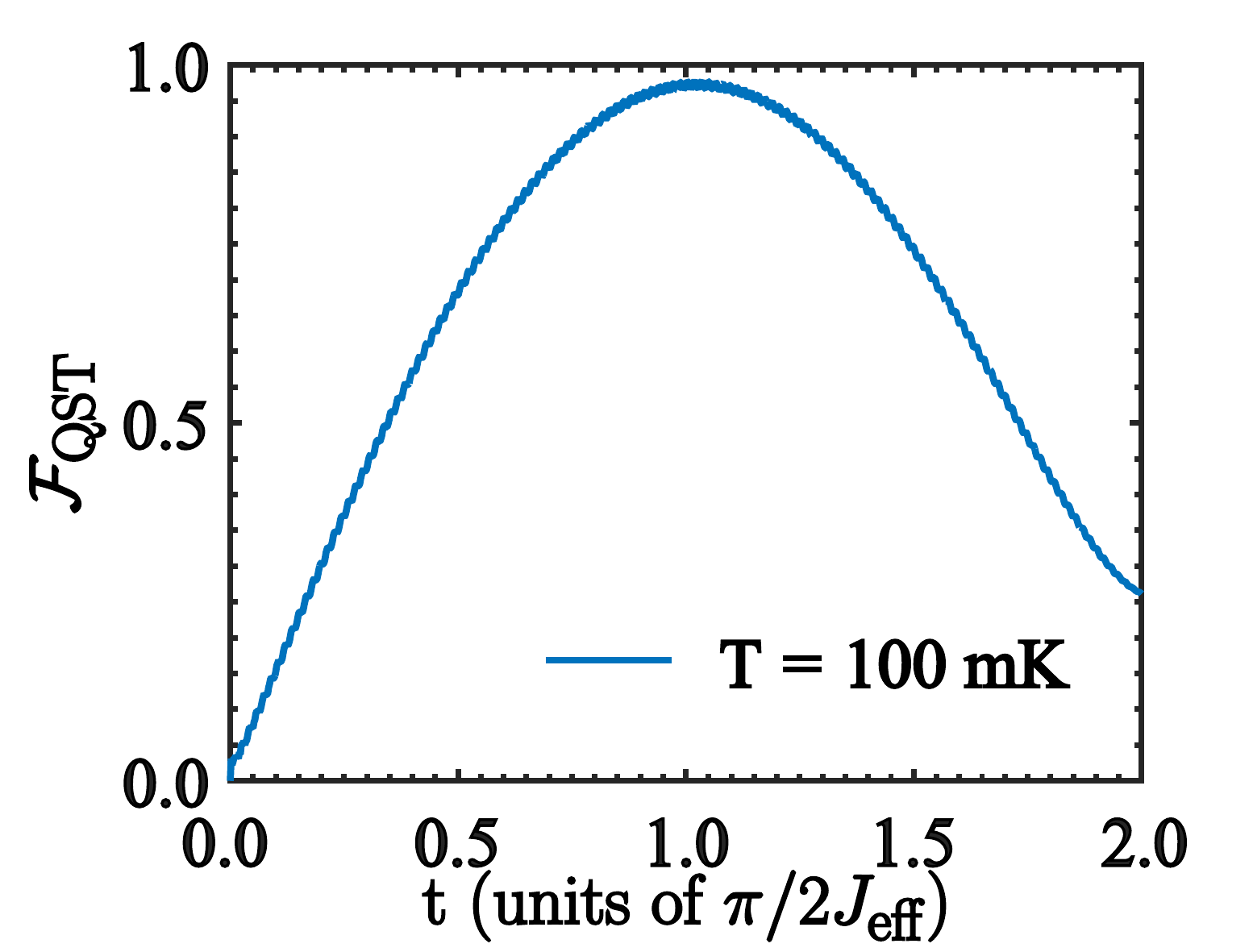}
	\caption{\textbf{Fidelity of the QST process}. Time-evolution of the QST fidelity $\mathcal{F}_{\rm QST}$ between the states $\ket{\chi}=\cos\theta\ket{\uparrow}+\sin\theta e^{i\phi}\ket{\downarrow}$ and $\rho_{Q_2}(t)$, averaged over the Bloch sphere. We have considered the QRS initialized in a thermal state at $T=100$mK.}
	\label{fig:dissipativeQST}
\end{figure}

For the purpose of QST of an arbitrary qubit state, we need to excite either the leftmost or rightmost qubit, see Fig. \ref{fig:model}. This can be done by applying an external driving on the leftmost (rightmost) qubit $H_{d}(t) = \hbar \Omega\cos(\nu t+\phi)\tau_{1}^x(\tau_{2}^x)$ resonant with the qubit frequency gap $\omega_1$ ($\omega_2$) and far off-resonance with both the QRS and rightmost (leftmost) qubit. For example, one can initialize the whole system in the state
\begin{ceqn}
\begin{eqnarray}
\label{initial}
\rho_{\rm{QST}} = \rho_{\rm{th}}^{\rm{\small{QRS}}} \otimes \ket{\chi}\bra{\chi} \otimes \ket{\downarrow}\bra{\downarrow},
\end{eqnarray}
\end{ceqn}
where $\ket{\chi}=\cos\theta\ket{\uparrow}+\sin\theta e^{i\phi}\ket{\downarrow}$.

We study the QST under dissipative mechanisms in a way that is consistent with a circuit QED implementation based on a flux qubit coupled to an on-chip microwave cavity in the USC regime \cite{Niemczyk2010}, and including the effect of a finite temperature. The treatment that we use has also been applied to a recent implementation of the QRM at finite temperature~\cite{Forn2016}. Moreover, the leftmost and rightmost qubits could be implemented by means of transmon qubits \cite{transmon,PhysRevLett.112.070502} coupled to the edges of the microwave cavity. The dissipative dynamics is governed by the microscopic master equation in the Lindblad form \cite{PhysRevA.84.043832}
\begin{ceqn}
\begin{eqnarray}
\label{mastereq}
\frac{d\rho(t)}{dt} &=& \frac{i}{\hbar}[\rho,H]+  \sum_{j,k>j}(\Gamma_{X}^{jk}+\Gamma_{\gamma}^{jk})\mathcal{D}[\ketbra{\psi_{j}}{\psi_{k}}]\rho + \sum_{j=1}^{N=2}\gamma_{j}\mathcal{D}[\tau_{j}^{x}]\rho + \sum_{j=1}^{N=2}\gamma_{\phi_{j}}\mathcal{D}[\tau_{j}^{z}]\rho,
\end{eqnarray}
\end{ceqn}
where the Hamiltonian $H$ is given by Eq.~(\ref{model}), and $\mathcal{D}[O]\rho=1/2(2O\rho O^{\dag} - \rho O^{\dag}O-O^{\dag}O\rho)$. Here, $\gamma_{j}$ corresponds to the relaxation rate for the external qubit and $\gamma_{\phi_{j}}$ stands for the qubit pure dephasing rate. Also, $\Gamma_{X}^{jk}$ is the dressed photon leakage rate for the resonator, $\Gamma_{\gamma}^{jk}$ corresponds to the dressed qubit relaxation rate for the qubit-cavity system, both dressed rates are defined as follows
\begin{ceqn}
\begin{eqnarray}
\Gamma_{X}^{jk} &&=\frac{\kappa}{\omega_{\rm cav}}\nu_{kj}|X_{jk}|^2\\
\Gamma_{\gamma}^{jk} &&=\frac{\gamma}{\omega_{q,i}}\nu_{kj}|\sigma^{x}_{jk}|^2,
\end{eqnarray}
\end{ceqn}
where $\kappa$ and $\gamma$ are the bare decay rates for the cavity mode and the two-level systems that belong to the QRS, $\nu_{kj}=\nu_k-\nu_j$, and $|X_{jk}|^2=|\bra{\psi_j}(a+a^{\dag})\ket{\psi_k}|^2$, $|\sigma^x_{jk}|^2=|\bra{\psi_j}\sigma^x_i\ket{\psi_k}|^2$. The QST fidelity $\mathcal{F}_{\rm QST}=\bra{\chi}\rho_{Q_2}(t)\ket{\chi}$, where $\rho_{Q_2}(t)$ is the density matrix of the rightmost qubit, is numerically calculated from Eq.~(\ref{mastereq}) for $4000$ input states $\ket{\chi}$, uniformly distributed over the Bloch sphere. Figure \ref{fig:dissipativeQST} shows the evolution of $\mathcal{F}_{\rm QST}$ as a function of time and for a system temperature of $T=100$~mK. As the QST time is shorter than any decay rate in the system, the dissipation should not affect the performance of QST. We see that the main detrimental effect on QST is produced by the distribution of the thermally populated states in the mediator. The numerical calculations have been performed for the QRS parameters given in Ref.~\cite{Forn2016}, that is, $\kappa/2\pi=0.10~$MHz, $\gamma/2\pi=15~$MHz, and for the leftmost and rightmost qubits we choose $\gamma_j/2\pi=0.48~$MHz, $\gamma_{\phi_j}/2\pi=0.15~$MHz given in Ref.~\cite{PhysRevLett.112.070502}. At temperature $T=100$~mK we obtain maximum fidelity of about $\mathcal{F}_{\rm QST}=0.9785$. It is worth mentioning that the validity of our QST protocol relies on the range of temperatures in which the system can be initialized as a product state $\rho_{\rm p}$.

\section*{Experimental Proposal}
\begin{figure}[t!]
\centering
\includegraphics[width=20cm,height=10cm,keepaspectratio]{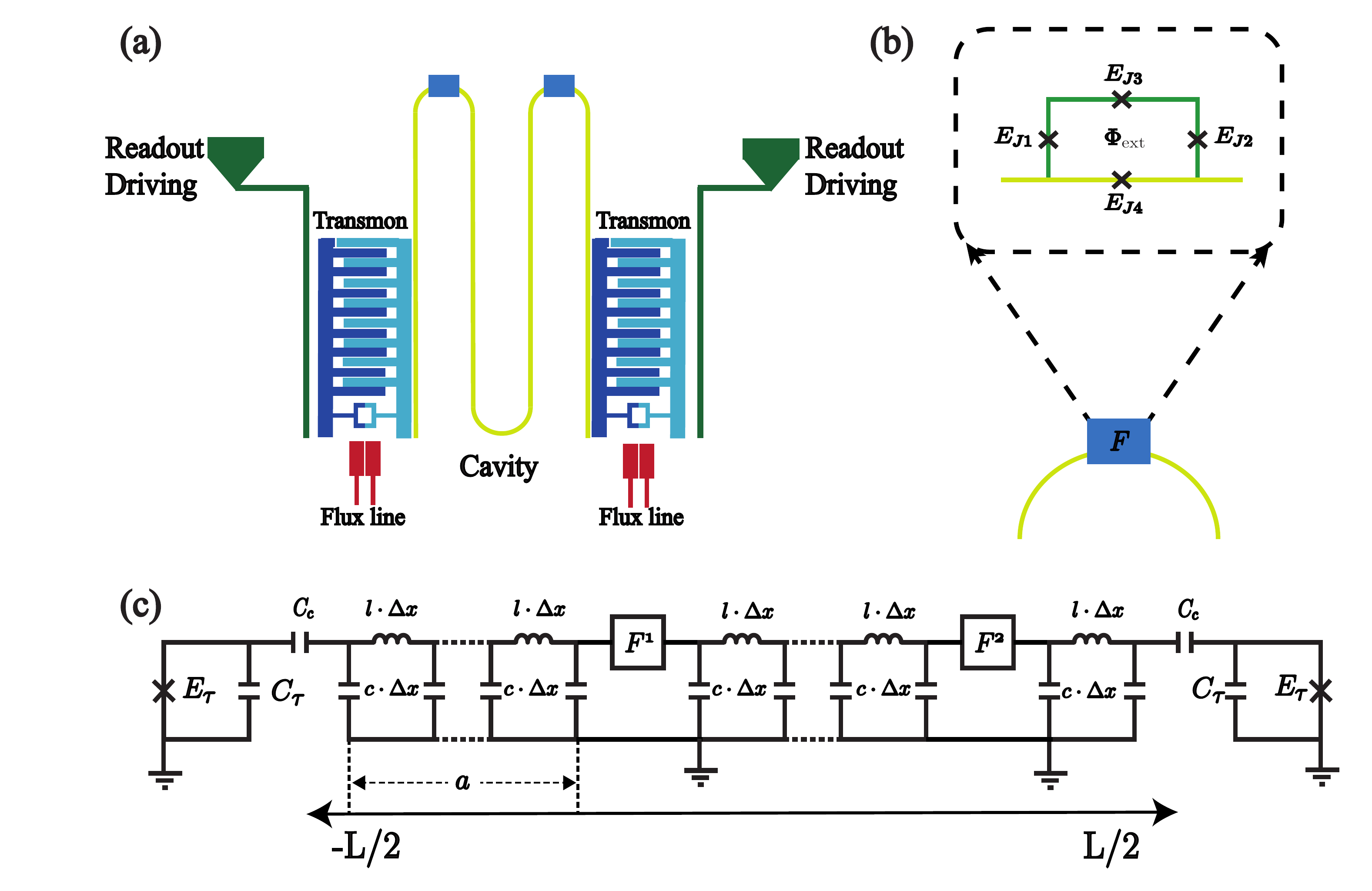}
\caption{\textbf{Schematic of the experimental proposal.} (a) A superconducting $\lambda/2$ coplanar waveguide resonator is galvanically coupled to $N=2$ superconducting loops formed by four Josephson junctions. Also, the resonator is coupled capacitively to two transmon devices at the edges of the waveguide. (b) Enlarged view of figure (a) at the position where the superconducting loop is placed. The flux qubit is formed by three Josephson junctions and the coupling between each flux qubit and the resonator is mediated by the the fourth embedded junction. (c) Equivalent circuit for (a), the resonator is considered as a finite set of LC circuits inductively connected in a series, these LC circuits are characterized by the capacitance $c\cdot\Delta x$ and inductance $l\cdot\Delta x$, the flux qubits are denoted by $F^{(j)}$ and the transmon devices are formed by two superconducting islands shunted by a large capacitance $C_{\tau}$ and an SQUID loop.}
\label{fig:circuit}
\end{figure}
Our proposal for quantum state transfer might be implemented on the circuit QED architecture shown in Fig. \ref{fig:circuit}(\textcolor{blue}{a}), where a $\lambda/2$ coplanar waveguide resonator (CPWR) is galvanically coupled to $N=2$ superconducting loops each formed by four Josephson junctions (JJs), see Fig. \ref{fig:circuit}(\textcolor{blue}{b}). Three junctions in the superconducting loop will form the flux qubit \cite{Phys.Rev.B.60.15398} while the fourth junction embedded into the CPWR, and characterized by the plasma frequency $\omega_p$, will play the role of coupling junction. Also, two transmon qubits \cite{transmon} are capacitively coupled to the resonator through the voltage at the ends of the CPWR. The equivalent circuit is shown in Fig. \ref{fig:circuit}(\textcolor{blue}{c}), where CPWR is modeled as an finite array of LC circuit inductively connected in a series, each lumped circuit element is characterized by the capacitance $c\cdot\Delta x$ and inductance $l\cdot\Delta x$, where $\Delta x$ stands for the lattice space. Moreover, the embedded Josephson junctions are located at the positions $x=a=L/(N+1)$, with $L$ the resonator length and $N=2$ the number of Josephson junctions that interrupt the coplanar waveguide. In this case, the state of the circuit can be characterized by the flux node $\phi(x,t) = \int_{-\infty}^{t}V(x,t')dt'$ where $V(x,t')$ is the voltage associated with in each system component with respect to the ground plane. The circuit Lagrangian reads

\begin{ceqn}
	\begin{eqnarray}
	\label{lagrangian}
	\mathcal{L} = \sum_{i=1}^{N}\mathcal{L}_{\rm{JJ}}^{(i)} + \sum_{i=1}^{N+1}\mathcal{L}_{\rm{Tl}}^{(i)}  + \sum_{j=1}^{2}\mathcal{L}_{\rm{Q}}^{(j)} + \mathcal{L}_{\rm{I}},
	\end{eqnarray}
\end{ceqn}
where
\begin{ceqn}
	\begin{eqnarray}
	\mathcal{L}_{\rm{JJ}}^{(i)} &&=\sum_{l=1}^{4} \frac{C_{J_{l}}^{(i)}}{2}(\dot{\phi}_{l}^{(i)})^2 + E_{J_{l}}^{(i)}\cos\bigg(\frac{\phi_{l}^{(i)}}{\varphi_{0}}\bigg) \\
	\mathcal{L}_{\rm{Tl}}^{(i)} &&= \sum_{k=0}^{n}\frac{c\cdot\Delta x}{2}(\dot{\psi}_{k}^{(i)})^2 - \sum_{k=0}^{n-1}\frac{1}{2l\cdot\Delta x}(\psi_{k}^{(i)}-\psi_{k+1}^{(i)})^2\\
	\mathcal{L}_{\rm{Q}}^{(j)} &&=\frac{C_{\tau}^{(j)}+C_{c}}{2}(\dot{\phi}_{\tau}^{(j)})^2 + E_{\tau,j}\cos\bigg(\frac{\phi_{\tau}^{(j)}}{\varphi_{0}}\bigg)\\
	\mathcal{L}_{\rm{I}} &&=\frac{C_{c}}{2}(\dot{\phi}_{\tau}^{1}-\dot{\psi}_{0}^{(1)})^2 +\frac{C_{c}}{2}(\dot{\phi}_{\tau}^{2}-\dot{\psi}_{n}^{(N+1)})^2.
	\end{eqnarray}
\end{ceqn}
Here $\psi_{k}^{(i)}$, $\phi_{l}^{(i)}$, and $\phi_{\tau}^{(j)}$ correspond to the flux variables that describe the CPWR, the junctions in the superconducting loops, and the transmon qubits, respectively. Furthermore, $C_{J_{l}}^{(i)}$ and $E_{J_{l}}^{(i)}$ are the Josephson capacitance and energy on each JJ that compounds the flux qubit and the coupling junction. $C_{\tau}^{(j)}$  are the total capacitances of the transmon qubits and $E_{\tau,j}$ their respective Josephson energies. $C_{c}$ is the coupling capacitance between the transmon qubit and the coplanar waveguide. Following the formal procedure for the circuit quantization, the Hamiltonian of the system reads
\begin{ceqn}
\label{Dicke}
\begin{eqnarray}
H &&= \sum_{r\in\mathcal{M}} \omega_{r}a_{r}^{\dag} a_{r} + \sum_{i}\frac{\omega_{q,i}}{2}\sigma^{z}_{i} + \sum_{r\in\mathcal{M},i}g_{r,i}\sigma^{x}_{i}(a_{r} + a_{r}^{\dag})\nonumber\\
\label{transmon}
&&+ \sum_{j}4E_{C,j}\sum_{n}\ketbra{n_j}{n_j} + \frac{E_{\tau,j}}{2}\sum_{n}\big(\ketbra{n_j}{n_j+1} + \ketbra{n_j+1}{n_j}\big)\nonumber\\
&&+  \sum_{r\in\mathcal{M},j}\lambda_{j,r}\ketbra{n_j}{n_j}(a_{r} + a_{r}^{\dag}).
\end{eqnarray}
\end{ceqn}

The detailed derivation of the above Hamiltonian can be found in the supplemental material. Here, $a_{r}$($a_{r}^{\dag}$) is the annihilation(creation) bosonic operator of the CPWR belonging to the manifold $\mathcal{M}$, $\sigma^{z,x}_{i}$ are the Pauli matrices that describe the two-level system formed by the flux qubit. In addition, $\omega_{q,i}$ and $\omega_{r}$ are the frequencies for the qubit and the field mode, respectively, and $g_{i,r}$ is the coupling strength between the CPWR and the two-level system. Also, $E_{C,j}$ is the charge energy of each transmon and $E_{\tau,j}$ its respective Josephson energy. Notice that the charge qubit Hamiltonian will be truncated up to the third energy level, $n=0,1,2$. This is because of in the transmon regime, $E_{C}\ll E_{J}$, the anhamonicity $\alpha=E_{10}-E_{21}$ is only $\sim~3-5~\%$ of $E_{10}$ \cite{Phys. Rev. A 82.040305}. However, our numerical results with realistic circuit QED parameters show that the third energy level does not take part in the protocol (see supplemental material) and the charge qubit can be considered as an effective two-level system. In this computational basis the interaction between the transmon and the field mode becomes proportional to $\tau_{j}^{x}(a_r^{\dag}+a_r)$ as in Eq.~(\ref{model}). 
\begin{figure}[t!]
\centering
\includegraphics[width=9cm,height=9cm,keepaspectratio]{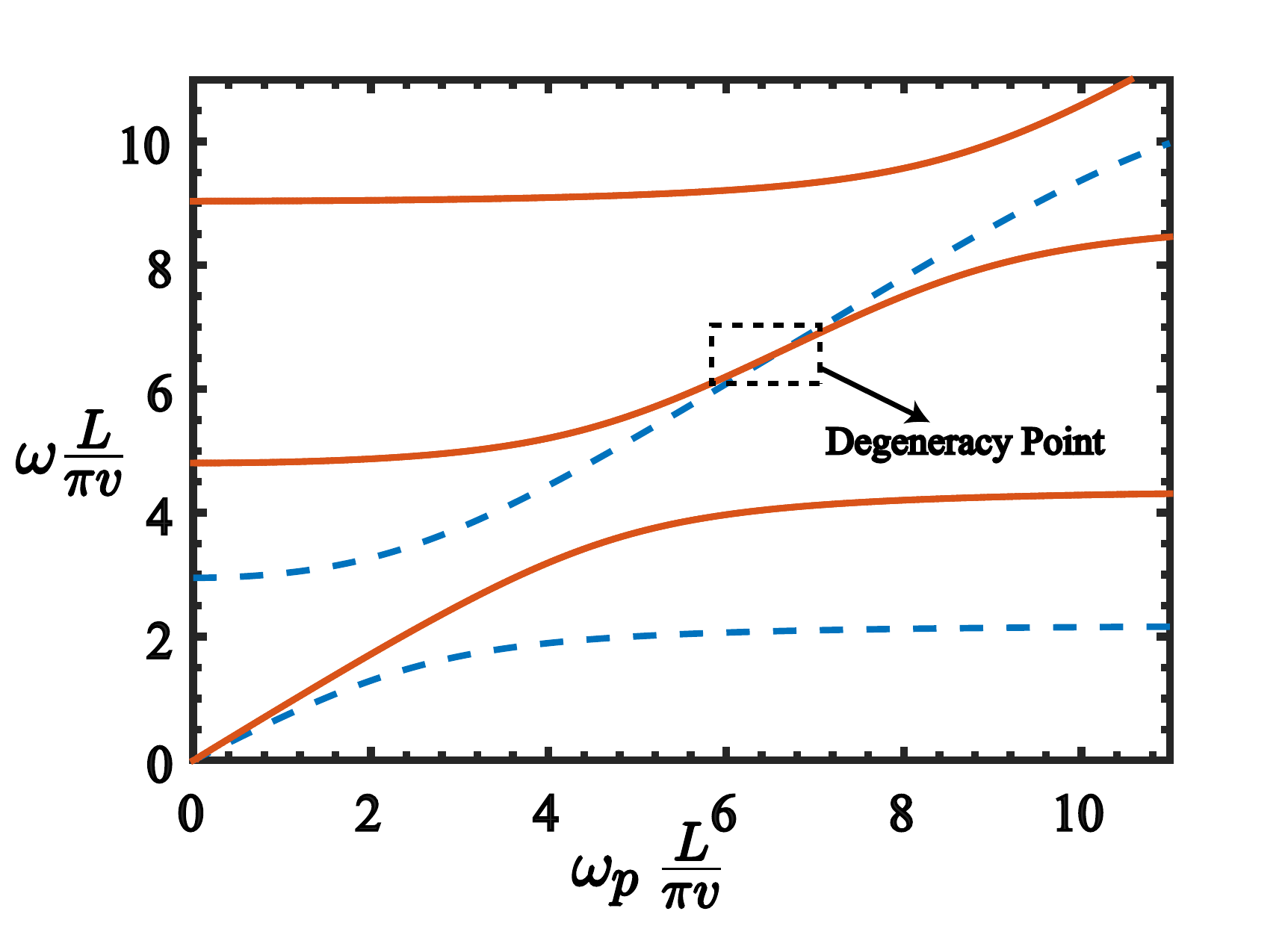}
\caption{\textbf{Energy spectrum of the CPWR.} Energy spectrum of the coplanar waveguide resonator with two Josephson junctions embedded on it, as a function of the plasma frequency $\omega_{p}$. The parameters for the resonator are $v=0.98\times10^{8}\rm{m/s}$, $Z_{0}=50\Omega$, $L=0.28\rm{mm}$, and the Josephson capacitance of the embedded junction is $C_{J_4}=1\rm{pF}$. For particular values of the plasma frequency, the eigenmodes belonging to an specific manifold become degenerate.}
\label{fig:manifold}
\end{figure}

On the other hand, each eigenmode frequency of the CPWR will constitute a manifold $\mathcal{M}$ which has as many modes as the number of embedded junctions. Nevertheless, for an specific value of the plasma frequencies of the embedded junctions, that is, $\omega_{p} = \pi v(N+1)/L$, the modes in the manifold $\mathcal{M}$ become degenerate (cf. Fig \ref{fig:manifold}), single-band approximation \cite{Phys.Rev.Lett.112.223603}, allowing to obtain the generalized Dicke model \cite{Phys.Rev.Lett.114. 143601}
\begin{ceqn}
\begin{eqnarray}
\label{newrabi}
H&&=  H'_{\rm{QRS}} + \sum_{r\in\mathcal{M},j}\frac{\omega_j}{2}\tau_{j}^{z} + \lambda_{j,r}\tau_{j}^{x}(a_{r}^{\dag}+a_{r})\\
H'_{\rm{QRS}} &&= \sum_{r\in\mathcal{M}} \omega_{r}a_{r}^{\dag} a_{r} + \sum_{i}\frac{\omega_{q,i}}{2}\sigma^{z}_{i} + \sum_{r\in\mathcal{M},i}g_{r,i}\sigma^{x}_{i}(a_{r} + a_{r}^{\dag}).
\end{eqnarray}
\end{ceqn}

\begin{figure}[t!]
\centering
\includegraphics[width=9cm,height=9cm,keepaspectratio]{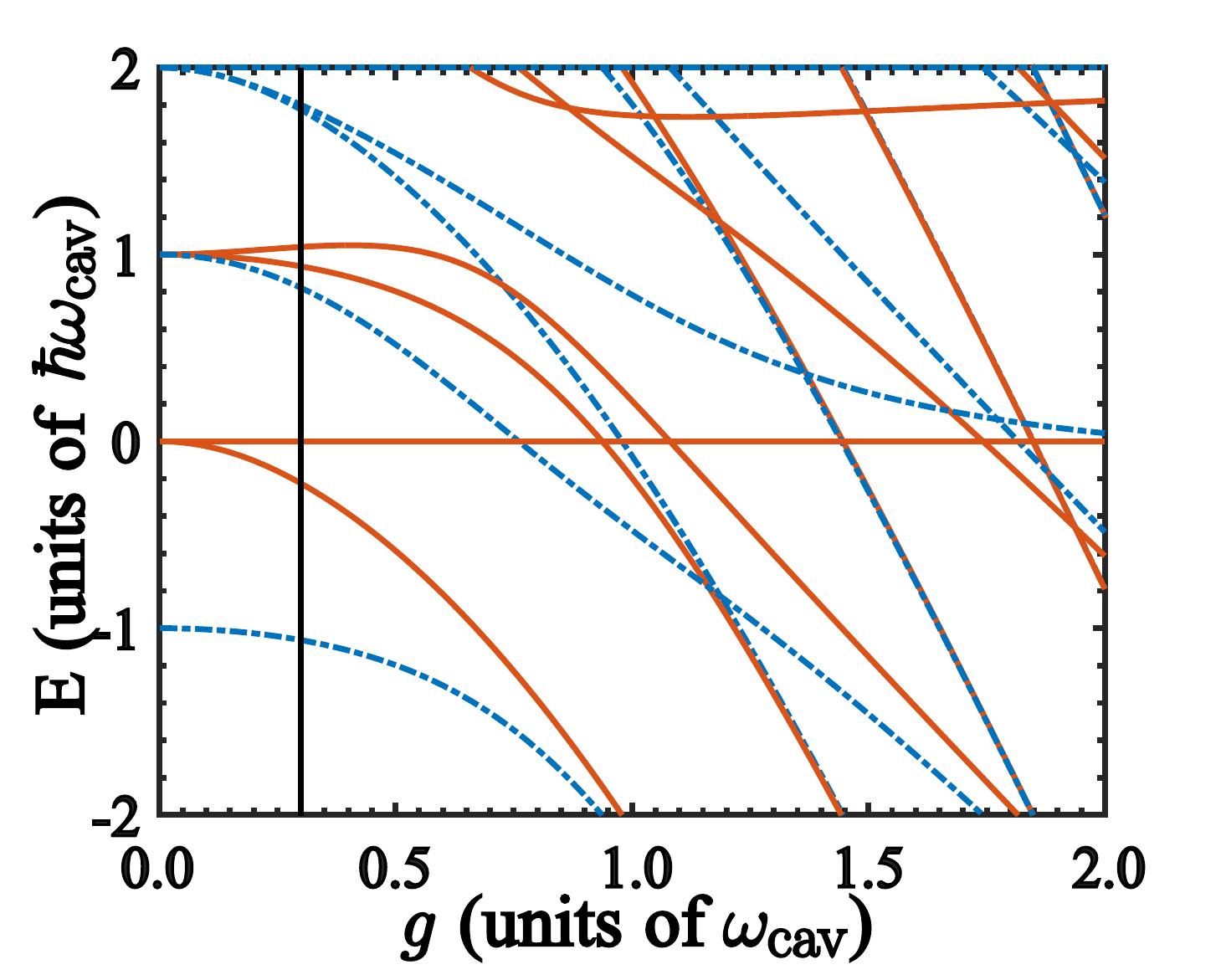}
\caption{\textbf{Energy spectrum of the generalize Dicke model.} Energy spectrum of the Hamiltonian (\ref{newrabi}) with parameters $\omega_{q,1} = \omega_{q,2} = \omega_{\rm{cav}}$, as a function of the coupling strength $g$. Blue (dot-dashed) lines stand for states with parity $p=+1$, and red (continuous) lines stand for states with parity $p=-1$. Straight lines stand for dark states.}
\label{fig:spectrum2}
\end{figure}

\begin{figure}[H]
	\centering
	\includegraphics[width=10cm,height=10cm,keepaspectratio]{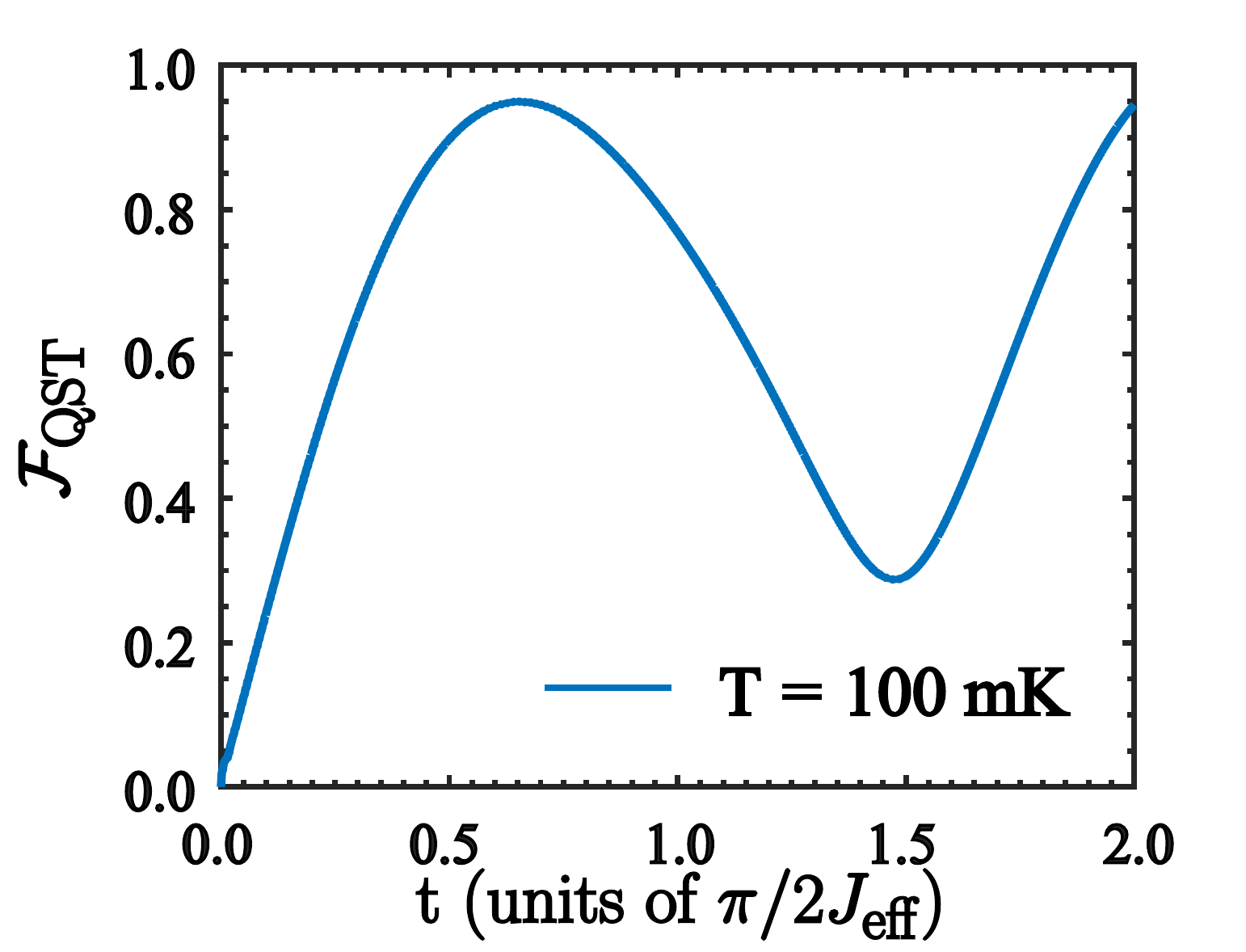}
	\caption{\textbf{Fidelity of the QST process for the generalized Dicke model.} Time-evolution of the QST fidelity $\mathcal{F}_{\rm QST}$ between the states $\ket{\chi}=\cos\theta\ket{\uparrow}+\sin\theta e^{i\phi}\ket{\downarrow}$ and $\rho_{Q_2}(t)$, averaged over the Bloch sphere. We have considered the generalized Dicke mediator initialized in a thermal state at $T=100$mK.}
	\label{fig:fidelity}
\end{figure}

The Hamiltonian $H'_{\rm{QRS}}$ also exhibits the $\mathbb{Z}_{2}$ symmetry as the Hamiltonian (\ref{QRS}). Therefore, the energy spectrum of  $H'_{\rm{QRS}}$ can be tagged in terms of two parity subspaces as shown in Figure \ref{fig:spectrum2}. Notice that the spectrum presents similar features compared with the spectrum of Figure \ref{fig:spectrum}, that is, both the QRS and the modified QRS described by  $H'_{\rm QRS}$ have the same energy level configuration in terms of parity subspaces and dark states. Under those circumstances, the generalized Dicke model in Eq. (\ref{newrabi}) and the model in Eq. (\ref{model}) present the same selection rules which are fundamental in our protocol for state transfer. In order to study how the protocol works for the generalized Dicke model, we start by obtaining the Gibss state of the Hamiltonian (\ref{newrabi}) at $T=100\rm{mK}$. To assure that the thermal state is a tensor product between the external qubits and the system mediator, we compute the fidelity between the Gibbs state and the probe state defined by $\rho_{\rm{p}}=\rho_{\rm{th}}^{\rm{\small{QRS}}}\otimes |\downarrow\downarrow\rangle\langle\downarrow\downarrow|$. If we consider realistic parameters for loss mechanisms discussed in the description the master equation, we have obtained a fidelity of $\mathcal{F}={\rm tr}(\rho_{\rm Gibbs}\rho_{\rm p})=0.9443$. The result of our transfer protocol is shown in Figure \ref{fig:fidelity}. As discussed before, we consider $4000$ different initial states for the leftmost qubit, $\ket{\chi}=\cos\theta\ket{\uparrow}+\sin\theta e^{i\phi}\ket{\downarrow}$, uniformly distributed over the Bloch sphere. The result is similar to the one obtained with the model described by Eq. (\ref{model}), that is, for the same physical parameters of loss mechanisms we obtain that at temperature $T=100\rm{mK}$ the fidelity reach its maximum $\mathcal{F}_{\rm{max}}=0.9503$. Higher temperatures in the physical setup will diminish this fidelity. 

\section*{Conclusion} 
We have shown that a system composed by two qubits connected to an incoherent QRS mediator, allows us to carry out high fidelity QST of single-qubit states even though the mediator system is in a thermally populated state. The QST mechanism involves the tuning of qubit frequencies resonant to a parity forbidden transition in the QRS such that an effective qubit-qubit interaction appears. Numerical simulations with realistic circuit QED parameters show that QST is successful for a broad range of temperatures. We have also discussed a possible physical implementation of our QST protocol for a realistic circuit QED scheme that leads to the generalized Dicke model. Our proposal may be of interest for hot quantum information processing within the context of ultrastrong coupling regime of light-matter interaction.

\section*{Author contributions} G.R and J.C.R supervised and contributed to the theoretical analysis. F.A.C.-L carried out all calculations. F.A-A prepared the figures and F.A.C.-L, G.A.B and G.R wrote the manuscript. All authors contributed to the results discussion and revised the manuscript.

\section*{Acknowledgements} F.A.C.-L acknowledges support from  CEDENNA basal grant No. FB0807 and Direcci{\'o}n de Postgrado USACH, F.A.-A acknowledges support CONICYT Doctorado Nacional 21140432, G.A.B acknowledges support from CONICYT Doctorado Nacional 21140587, J. C. R acknowledges the support from FONDECYT under grant No. 1140194, and G.R acknowledges the support from FONDECYT under grant No. 1150653.

\section*{Additional information} Competing financial interests: The authors declare no competing financial interests.

\end{document}